\documentclass[prd,eqsecnum,preprintnumbers,showpacs]{revtex4}
\usepackage{amsfonts}
\usepackage{amsmath}
\usepackage{amssymb}

\begin{document}
\title[Gravitational-wave probe of effective quantum gravity]{A gravitational-wave probe of effective quantum gravity}

\author{Stephon Alexander}
\altaffiliation{Center for Fundamental Theory, Institute for
Gravitational Physics and Geometry, Department of Physics}
\affiliation{The Pennsylvania State University, University Park, PA 16802, USA}
\author{Lee Samuel Finn}
\altaffiliation{Center for Gravitational Wave Physics, 
Institute for Gravitational Physics and Geometry, 
Department of Physics, Department of Astronomy and Astrophysics}
\affiliation{The Pennsylvania State University, University Park, PA 16802, USA}
\author{Nicol\'as Yunes}
\altaffiliation{Center for Gravitational Wave Physics, 
Institute for Gravitational Physics and Geometry, 
Department of Physics}
\affiliation{The Pennsylvania State University, University Park, PA 16802, USA}

\date{\today}

\begin{abstract}
  
  The Green-Schwarz anomaly-cancelling mechanism in string theories
  requires a Chern-Simons term in the Einstein-Hilbert action, which
  leads to an amplitude birefringence of spacetime for the propagation
  of gravitational waves.
While the degree of birefringence may be intrinsically small, its effects on a gravitational wave will accumulate as the wave propagates. 
The proposed Laser Interferometer Space Antenna (LISA) will be sensitive enough to observe the gravitational waves from sources at cosmological distances great enough that interesting bounds on the Chern-Simons may be found. Here we evaluate the effect of a Chern-Simons induced spacetime birefringence to the propagation of gravitational waves from such systems. 
We find that gravitational waves from in coalescing binary black hole system are imprinted with a signature of Chern-Simons gravity. This signature appears as a time-dependent change in the apparent orientation of the binary's orbital angular momentum with respect to the observer line-of-sight, with the change magnitude reflecting the integrated history of the Chern-Simons coupling over the worldline of a radiation wavefront. 
While spin-orbit coupling in the binary system will also lead to an evolution of the system's orbital angular momentum, the time dependence and other details of this \emph{real} effect are different than the \emph{apparent} effect produced by Chern-Simons birefringence, allowing the two effects to be separately indentified. 
In this way gravitational wave observations with LISA may thus provide our first and only opportunity to probe the quantum structure of spacetime over cosmological distances. 
  
\end{abstract}
{\small\catcode`\$=12{}%
\preprint{$Id: ms.tex 603 2007-12-15 22:24:47Z lsf $}
\catcode`\$=3}

\pacs{11.25.Wx, 95.55.Ym, 04.60.-m, 04.80.Cc}

\maketitle

\section{Introduction}\label{intro}

``Gravitational wave'' is the name we give to a short-wavelength
feature of the structure of spacetime, the arena within which all other
phenomena play-out their roles. As such, the direct observation of
gravitational waves offers an unprecedented opportunity to explore the 
environment that both enables and constrains the action of the broader
phenomena of nature.  Here we describe an observational
probe of quantum gravity made possible by the forthcoming
generation of space-based gravitational wave detectors.

In most corners of the perturbative string theory moduli space
({\emph{i.e.~}}Type IIB, I, Heterotic) in four-dimensional
compactifications, the addition of a parity-violating Chern-Simons
term to the Einstein-Hilbert is required for mathematical
consistency~\footnote{In the case of string theory, the absence of a
  Chern-Simons term leads to the Green-Schwarz anomaly. Quantum
  consistency requires cancellation of this anomaly. In order to
  eliminate the anomaly, the introduction of a Chern-Simons term is
  essential.  Heterotic M-theory makes use of an anomaly inflow, which
  also leads to the same requirement.}. Furthermore in the presence of
the Ramond-Ramond scalar (D-instanton charge), the Chern-Simons term
is induced in {\emph{all}} string theories due to duality
symmetries~\cite{polchinski:1998:stv,alexander:2006:css}.
The Chern-Simons term is not unique to string theory, but in fact it
could also arise in Loop Quantum Gravity, where the coupling is not
necessarily limited to small values. In the strong gravity sector of
this framework, this term arises to ensure invariance under large
gauge transformations of the Ashtekar connection
variables~\cite{ashtekar:1989:cpi}.
This correction plays the same role in string theory that an analogous
anomaly-canceling correction to the quantum-chromodynamics path
integral plays in the Standard Model \cite{alvarez-gaume:1984:ga}. While the anomaly-cancelling field in the Standard Model case interacts with photons, leading to significant observational constraints on its coupling, the anomaly canceling term considered here only affects the gravitational sector of the theory and is mostly unconstrained by observation \cite{alexander:2007:npp,alexander:2007:ppe,smith:2007:eoc}.  

The Chern-Simons correction to the Einstein-Hilbert action,
Chern-Simons gravity~\cite{jackiw:2003:cmo}, has been recently studied
in connection with tests both in the
cosmological~\cite{lue:1999:cso,li:2007:ccv,alexander:2006:icp,alexander:2006:css}
and the weak-gravitational
contexts~\cite{alexander:2007:npp,alexander:2007:ppe,smith:2007:eoc}.
In the context of gravitational wave theory, Chern-Simons gravity
leads to an \emph{amplitude birefringence} of spacetime for
gravitational wave propagation~\cite{jackiw:2003:cmo,alexander:2006:css,alexander:2005:bgw}: i.e., a polarization dependent amplification/attenuation of wave amplitude with distance propagated. Observation of gravitational waves that have propagated over cosmological distances, such as will be possible with the Laser Interferometer Gravitational Wave Observatory (LISA)  \cite{merkowitz:2007:lis,baker:2007:lpt,nrc:2007:nbe}, provide the opportunity to measure or bound the magnitude of the birefringence and, correspondingly, provide the first experimental constraints on string theory models of gravity. 

Gravitational wave observations have long been recognized as a tool
for testing our understanding of gravity (see~\cite{will:2006:cbg} for
a recent review).  Eardley and collaborators
\cite{eardley:1973:goa,eardley:1973:goa:1} first proposed a far-field
test of all metric theories of gravity through gravitational wave
observations.  Finn \cite{finn:1985:gwf}, and later Cutler and
Lindblom~\cite{cutler:1996:gh}, proposed a means of realizing these
measurements using a space-based detector in a circumsolar orbit
observing solar oscillations in the far-zone field.
Ryan~\cite{ryan:1997:aoe} argued that observations of the phase
evolution of the gravitational waves emitted during the
gravitational-wave driven inspiral of, e.g., a neutron star or stellar
mass black hole into a supermassive black hole could be used to
``map-out'' the spacetime metric in the vicinity of the black hole
horizon, testing the predictions of general relativity in the regime
of strong-fields. There have been several proposals describing
different ways in which gravitational wave observations could be used
to place bounds on the graviton Compton-wavelength \cite{%
  will:1998:bmo,%
  finn:2002:bmo,%
  sutton:2002:bgm,%
  will:2004:tat,%
  berti:2005:tgr,%
  berti:2005:esb}, the existence of a scalar component to the
gravitational interaction \cite{%
  scharre:2002:tsg,%
  will:2004:tat,%
  berti:2005:tgr,%
  berti:2005:esb}, and the existence of other corrections to general
relativity as manifest in some fundamental, dimensionful length scale
\cite{dreyer:2004:bhs,berti:2006:gso}.  The measurements we propose
here are, we believe, the first example of a direct model-independent
probe of string theory and quantum gravity with gravitational waves.

In section \ref{ref:CS} we review Chern-Simons modified gravity,
focusing attention on the scale of the Chern-Simons term and its
effect on the propagation of gravitational waves in a cosmological
background. In section \ref{sec:obs} we evaluate the observational
consequences of the Chern-Simons term in the context of ground- and
space-based gravitational wave detectors. In section
\ref{sec:conclusions} we summarize our conclusions and discuss avenues
of future research.

Conventions used in relativity work and conventions used in quantum field theory work are often at odds. We follow the relativity conventions Misner, Thorne and Wheeler \cite{misner:1973:g} in this work: in particular, 
\begin{itemize}
\item Our metric has signature ${-}+++{}$;
\item We label indices on spacetime tensors with greek characters and use latin indices to label indices on tensors defined on spacelike slices; 
\item We use a semicolon in an index list to denote a covariant derivative (i.e., $\nabla_{\mathbf{V}}\mathbf{U}$ becomes $V^\nu U_{\mu;\nu}$) and a comma to denote ordinary partial derivatives; 
\item Except where explicitly noted we work in \emph{geometric units},
  wherein $G=c=1$ for Newtonian gravitational constant $G$ speed of
  light $c$. 
\end{itemize}
Note that in geometric units, units of mass and length are interchangeable (i.e., $G/c^2$ has units of (length)/(mass)). This is in contrast to Planck units ($\hbar=c=1$), where units of mass and units of inverse length are interchangeable (i.e., $\hbar/c$ has units of $\textrm{(mass)}\times\textrm{(length)}$). 

\section{Chern-Simons Modified Gravity}
\label{ref:CS}

\subsection{Brief Review}

In this subsection we review the modification to classical General Relativity by the inclusion of a Chern-Simons term, based on~\cite{jackiw:2003:cmo,alexander:2005:bgw}. All four dimensional compactifications of string theory lead, via the
Green-Schwarz anomaly canceling mechanism, to the presence of a four
dimensional gravitational Chern-Simons term \cite{polchinski:1998:stv}. Chern-Simons forms are formally defined for odd dimensions, with the 3-form of particular interest for gauge theories. By introducing an embedding coordinate, which may be dynamical, Jackiw and Yi \cite{jackiw:2003:cmo} described a Chern-Simons correction to the Einstein-Hilbert action:
\begin{equation}
\label{eq:CS-action}
S_{\text{CS}} = \frac{1}{64\pi} \int d^4 x\, \theta R\,{}^*R  
\end{equation}
where $\theta$ is (a functional of) the embedding coordinate,
\begin{equation}
R\,{}^*R = \frac{1}{2}
R_{\alpha\beta\gamma\delta}\epsilon^{\alpha\beta\mu\nu}
{R^{\gamma\delta}}_{\mu\nu}, 
\end{equation}
and $\epsilon^{\alpha\beta\gamma\delta}$ is the Levi-Civita tensor
density. The variation with respect to the metric of this contribution
to the total action (which includes the Einstein-Hilbert action plus
the action corresponding to any additional matter fields) yields
\cite{jackiw:2003:cmo}
\begin{equation}
  \delta S_{\text{CS}} = -\frac{1}{16 \pi} \int d^4x \sqrt{-g} \;
  C_{\alpha\beta} \; \delta 
  g^{\alpha\beta}, 
\end{equation}
where $g$ is the determinant of the metric and $C^{ab}$ is the C-tensor~\footnote{In \cite{jackiw:2003:cmo}, $C_{ab}$ was defined as a
  $4$-dimensional generalization of the $3$-dimensional Cotton
  tensor. However, such a generalization already
  exists~\cite{garcia:2004:cti} and differs from that introduced in
  \cite{jackiw:2003:cmo}. Therefore, we shall here refer to this
  quantity as the C-tensor.}, 
\begin{equation}
  C^{\alpha\beta} = -\frac{1}{\sqrt{-g}} \left[ \theta_{,\epsilon}
    \epsilon^{\epsilon\gamma\delta(\alpha} \nabla_{\gamma} 
    R^{\beta)}{}_{\delta} - 
    \nabla_{\delta}\theta_{,\gamma} \;
    \epsilon^{\mu\nu\gamma(\alpha}{R^{\beta)\delta}}_{\mu\nu}
  \right],
\end{equation}
and the parenthesis in the superscript stand for symmetrization.  The
variation of $S_{\text{CS}}$, the usual Einstein-Hilbert action, and
the action of other matter fields leads to the equations of motion of
Chern-Simons modified gravity:
\begin{equation}
\label{eq:EOM}
G_{\alpha\beta} + C_{\alpha\beta} = 8 \pi T_{\alpha\beta},
\end{equation}
where $G_{ab}$ is the Einstein tensor (i.e., the trace-reversed Ricci
tensor) and $T_{ab}$ is the stress-energy tensor of the matter fields.

By construction the divergence of the Einstein tensor $G_{\alpha\beta}$ vanishes. If $\theta$ is treated as a fixed, external quantity then general covariance, which requires $\nabla\cdot\mathbf{T}=0$, leads to the constraint $\nabla\cdot\mathbf{C}=0$, which is shown in \cite{jackiw:2003:cmo} to be equivalent to $R\,{}^*R=0$.  Alternatively, if $\theta$ is a dynamical field, then variation of the action with respect to $\theta$ will lead to the same constraint on $R\,{}^*R$. Here we are interested in the propagation of gravitational waves in vacuum, where $\mathbf{T}=0$ and the constraint $\nabla\cdot\mathbf{C}=0$ is satisfied regardless of whether we view $\theta$ as a dynamical field or a fixed, externally-specified quantity. 

\subsection{Linearized Chern-Simons gravitational waves}
\label{LCS}

Focus attention on gravitational wave perturbations to a Friedmann-Robertson-Walker (FRW) cosmological background in Chern-Simons gravity. Following \cite{alexander:2005:bgw}, we can write the perturbed FRW line element as
\begin{equation}
\label{FRW}
ds^2 = a^2(\eta) \left[- d\eta^2 + \left(\delta_{ij} + h_{ij}\right) 
d\chi^i d\chi^j\right]
\end{equation}
where $\eta$ is conformal time, $\chi^i$ are comoving spatial
coordinates, $\delta_{ij}$ is the Euclidean metric, and $h_{ij}$ is
the metric perturbation, which --- for gravitational wave solutions
--- we can take to be transverse and traceless \footnote{One can show
  that a transverse-traceless gauge exists in Chern-Simons gravity, following
  the same steps as in classical general
  relativity~\cite{misner:1973:g}}.  Introducing this perturbation
into the field equations [Eq.~\eqref{eq:EOM}] leads to
\begin{equation}
\label{eq:h-eom}
\square_g h^{j}{}_{i} = - \frac{1}{a^2} \epsilon^{pjk} \left[ \left(
    \theta'' - 2 {\mathcal{H}} \theta' \right) \partial_ph'_{ki} + \theta' \;
    \partial_p\square_g h_{ki} \right],   
\end{equation}
where we have introduced the notation
\begin{eqnarray}
' &=& \partial_\eta\\ 
\square_g &=& \partial_{\eta}^2 - \partial_i \partial_j \gamma^{ij} +
2 {\mathcal{H}} \partial_{\eta} \\ 
\mathcal{H}&=& a'/a. 
\end{eqnarray}
Conformal time $\eta$ is related to proper time measured by an observer at
rest with respect to the cosmological fluid via
\begin{equation}
dt = a(\eta)d\eta;
\end{equation}
correspondingly, the conformal Hubble function $\mathcal{H}$ is related 
to the Hubble function $H$ measured by an observer at rest with
respect to slices of homogeneity via 
\begin{equation}
H \equiv \frac{\dot{a}}{a} = \frac{1}{a}\mathcal{H},
\end{equation}
where we have use dots to stand for derivatives with respect to cosmic
time $t$.

Focus attention on plane-wave solutions to the wave equation
[Eq.~\eqref{eq:h-eom}].  With the ansatz
\begin{equation}\label{eq:ansatz}
h_{lm}(\eta,\chi^l) 
= \frac{\mathcal{A}_{lm}}{a(\eta)}
e^{-i\left[\phi(\eta) - \kappa n_k\chi^k\right]},
\end{equation}
where the amplitude $\mathcal{A}_{lm}$, the unit vector in the
direction of wave propagation $n_k$ and the conformal wavenumber
$\kappa>0$ are all constant, we find that $\phi$, $\kappa$ and
$\mathcal{A}_{ij}$ must satisfy
\begin{equation}\label{eq:oka}
\mathcal{D}\mathcal{A}_{ij}= 
-a^{-2}\epsilon^{pjk}n_p\mathcal{A}_{ki}\left[
\left(\theta''-2\mathcal{H}\theta'\right)
\left(\phi'-i\mathcal{H}\right)\kappa 
+ i\theta' \kappa \mathcal{D}\right]
\end{equation}
where 
\begin{equation}
\mathcal{D} = -i\phi''-\left(\phi'\right)^2 -
\mathcal{H}'-\mathcal{H}^2+\kappa^2.
\end{equation}

Since the Chern-Simons correction breaks parity, it is convenient to resolve
$\mathcal{A}_{ij}$ into definite parity states, corresponding to
radiation amplitude in the right- and left-handed polarizations 
$e^{R}_{ij}$ and $e^{L}_{ij}$:
\begin{subequations}
\begin{eqnarray}
\mathcal{A}_{ij} &=& \mathcal{A}_{R}e^{R}_{ij} + \mathcal{A}_{L}e^{L}_{ij}
\end{eqnarray}
where 
\begin{eqnarray}
e_{kl}^{R} &=& \frac{1}{\sqrt{2}}\left(e^{+}_{kl} + ie^{\times}_{kl}\right)\\
e_{kl}^{L} &=& \frac{1}{\sqrt{2}}\left(e^{+}_{kl} - ie^{\times}_{kl}\right)
\end{eqnarray}
and $e^{+,\times}_{kl}$ are the usual linear polarization
tensors~\cite{misner:1973:g}.
\end{subequations}
It is straightforward to show that 
\begin{subequations}
\begin{equation}
\label{eigen}
n_i\epsilon^{ijk} e_{kl}^{\text{R,L}} = i
\lambda_{\text{R,L}} \left(e^j{}_l\right)^{\text{R,L}}, 
\end{equation}
where
\begin{eqnarray}
\lambda_{\text{R}} &=& +1\\
\lambda_{\text{L}} &=& - 1.
\end{eqnarray}
\end{subequations}
With this substitution equation~\eqref{eq:oka} becomes two 
decoupled equations, one for right-hand polarized waves and one for 
left-hand polarized waves:
\begin{equation}\label{eq:disp}
i\phi''_{\text{R,L}}+\left(\phi'_{\text{R,L}}\right)^2
+\mathcal{H}'+\mathcal{H}^2-\kappa^2=
\frac{i\lambda_{\text{R,L}}
\left(\theta'' - 2\mathcal{H} \theta'\right)
\left(\phi'-i\mathcal{H}\right)\kappa/a^2}%
{\left(1-\lambda_{\text{R,L}}\kappa \theta'/a^2\right)}
\end{equation}

The terms on the right-hand side of equation~\eqref{eq:disp} are the
Chern-Simons corrections to gravitational plane-wave propagation in a
FRW spacetime. To understand the relative scale of these terms, we
rewrite the equation in terms of the ratio $\phi'/\kappa$:
\begin{subequations}
\begin{eqnarray}\label{eq:disp-nodim}
\frac{y'}{\kappa} + 
i\left(1- \gamma^2\Gamma^2-\delta\Delta - y^2 \right) 
&=&
\frac{\lambda_{\text{R,L}}
\left(\epsilon E - 2\gamma\zeta\Gamma Z\right)}%
{1-\lambda_{\text{R,L}}\zeta Z}
\left(y-i\gamma\Gamma\right)
\end{eqnarray}
where 
\begin{eqnarray}
y &=& \frac{\phi'}{\kappa}\\
\gamma = \frac{\mathcal{H}_0}{\kappa} \qquad&\text{and}&\qquad
\Gamma = \frac{\mathcal{H}}{\mathcal{H}_0}\\
\delta = \frac{\mathcal{H}'_0}{\kappa^2} \qquad&\text{and}&\qquad
\Delta = \frac{\mathcal{H}'}{\mathcal{H}'_0}\\
\epsilon = \frac{\theta''_0}{a_0^2} \qquad&\text{and}&\qquad
E = \frac{\theta''}{a^2\epsilon}\\
\zeta = \frac{\kappa \theta'_0}{a^2_0} \qquad&\text{and}&\qquad
Z = \frac{\kappa \theta'}{a^2\zeta}
\end{eqnarray}
\end{subequations}
and a subscript $0$ indicates the present-day value of the functions
$\theta'$, $\theta''$, $\mathcal{H}$, $\mathcal{H'}$ and $a$.

If we assume that $\theta$ and $\mathcal{H}$ evolve on cosmological timescales (i.e., $f'\sim\mathcal{H}f$) then 
\begin{equation}
\epsilon^2 \sim \left(\gamma\zeta\right)^2 
\ll \gamma^2 \sim \left|\delta\right|.
\end{equation}

Treating the terms in $\epsilon$ and $\gamma\zeta$ as perturbations,
write the solution to equation~\eqref{eq:disp-nodim} as
\begin{equation}
y = y_0 + \epsilon y_{0,1} + \gamma\zeta y_{1,0} + \ldots
\end{equation}
where $y_0 = \phi_0'/\kappa$ is the solution to the unperturbed
equation [i.e., the dispersion relation in an FRW cosmology, given by
equation~\eqref{eq:disp-nodim} with vanishing right-hand side]. The first
corrections $y_{0,1}$ and $y_{1,0}$ owing to the Chern-Simons terms satisfy
\begin{subequations}
\begin{eqnarray}
y'_{0,1} - 2i\kappa y_0y_{0,1} &=& 
\lambda_{\text{R,L}} \kappa E y_0\\
y'_{1,0} - 2i\kappa y_0y_{1,0} &=& 
-2\lambda_{\text{R,L}} \kappa\Gamma Z y_0. 
\end{eqnarray}
\end{subequations}
Requiring that the perturbation vanish at some initial (conformal) time
$\eta_i$ the perturbations $y_{0,1}$ and $y_{1,0}$ satisfy
\begin{subequations}\label{eq:defY}
\begin{eqnarray}
y_{0,1}(\eta) &=& \lambda_{\text{R,L}}\mathcal{Y}[E](\eta)\\
y_{1,0}(\eta) &=& -2\lambda_{\text{R,L}} \mathcal{Y}[\Gamma Z](\eta)
\end{eqnarray} 
where
\begin{equation}
\mathcal{Y}[g](\eta)=
\kappa e^{2i\phi_0(\eta)}\int^\eta_{\eta_i}\!dx\, 
e^{-2i\phi_0(x)} y_0(x)g(x).
\end{equation}
\end{subequations}
Finally, the Chern-Simons correction to the accumulated phase as the 
plane-wave propagates from $\eta_i$ to $\eta$ is
\begin{equation}
\delta\phi_{\text{R,L}} = \kappa\lambda_{\text{R,L}}
\int_{\eta_i}^\eta d\eta\, 
\left\{
\epsilon\,\mathcal{Y}[E](\eta) -
2\gamma\zeta\,\mathcal{Y}[\Gamma Z](\eta)
\right\}
\end{equation}

When $\gamma\ll1$, i.e., $k_0$ is very much greater than the Hubble
constant ${H}_0$, the rescaled frequency $|y_0|\sim1$. In this
limit we can use integration by parts to find an asymptotic expansion
for $\mathcal{Y}[g]$:
\begin{eqnarray}
\mathcal{Y}[g](\eta) 
&\sim&
\frac{ie^{2i\phi_0(\eta)}}{2}\left[
e^{-2i\phi_0}\sum_{\ell=0}^n
\left(\frac{1}{2i\kappa}\right)^{\ell}
\left(\frac{1}{y_0}\frac{d}{d\eta}\right)^{\ell}g
\right]_{\eta_i}^\eta
+\mathcal{O}\left(\frac{1}{2i\kappa}\right)^{n+1}
\label{eq:asympt}
\end{eqnarray}

In the next section we explore the observational consequences of gravitational wave propagation in Chern-Simons gravity. 

\section{Observational consequences}\label{sec:obs}
\subsection{Birefringence in a Matter-Dominated Cosmology}

Current and proposed ground-based gravitational wave detectors are sensitive to gravitational waves in the $10$~Hz -- $1$~KHz band \cite{waldman:2006:sol,acernese:2006:vs,willke:2006:gp,luck:2006:sog}. Detectable sources in this band are expected to have redshifts $z\lesssim 1$. Space-based gravitational wave detectors like LISA \cite{merkowitz:2007:lis} will be sensitive to gravitational waves in the $0.1$ -- $100$~mHz band and, in this band, be sensitive enough to observe the gravitational waves from the inspiral of several $\sim10^6\,\text{M}_\odot$ black hole binary systems at $z\lesssim30$: i.e., anywhere in the universe they are expected \cite{bender:2007:lsr,baker:2007:lpu}. For sources in the band of these detectors 
\begin{equation}
\gamma = 3.7\times10^{-19}
\left(\frac{h_{100}}{0.72}\right)
\left(\frac{1\,\text{Hz}}{kc/2\pi}\right)\ll 1
\end{equation}
where
\begin{eqnarray}
h_{100} &=& \frac{H_0}{100\;\textrm{km s}^{-1}\textrm{Mpc}^{-1}}.
\end{eqnarray}
Additionally, for redshitfs $z\lesssim30$ the universe is well described by a matter-dominated FRW
cosmological model. In this section we evaluate the effect that the
Chern-Simons corrections described above have on propagation of
gravitational plane-waves through a matter-dominated FRW model.

In a matter-dominated FRW model the scale factor $a(t)$ satisfies \cite{misner:1973:g}
\begin{subequations}\label{eq:mdfrw}
\begin{equation}
\frac{a(\eta)}{a_0} = \eta^2 = \frac{1}{1+z}
\end{equation}
where, by convention, $\eta=1$ at the present epoch. In this model
and with this convention
\begin{eqnarray}
\mathcal{H} = \frac{2}{\eta} = 2\sqrt{1+z} \qquad &{\textrm{and}}& \qquad
\mathcal{H}_0 = 2, \\*
\gamma = \frac{2}{\kappa} \qquad& {\textrm{and}} &\qquad
\Gamma = \eta^{-1}= \sqrt{1+z},\\*
\delta = -\frac{2}{\kappa^2} \qquad& \textrm{and}& \qquad \Delta =
\eta^{-2} = 1+z, \\*
\epsilon = \frac{H_0^2}{4} \theta_0'' \qquad& \textrm{and}& \qquad E =
\frac{1}{\eta^4}\frac{\theta''}{\theta''_0}, \\*
\zeta =  \frac{H_0 k}{2} \; \theta_0'  \qquad& \textrm{and}& \qquad Z =
\frac{1}{\eta^4}\frac{\theta'}{\theta'_0}, \\*
\end{eqnarray}
Additionally, 
\begin{eqnarray}
E &=& \frac{1}{\eta^4}\frac{\theta''}{\theta''_0}\\*
Z &=& \frac{1}{\eta^4}\frac{\theta'}{\theta'_0}\\*
a_0 &=& \frac{\mathcal{H}_0}{H_0} = \frac{2}{H_0}.
\end{eqnarray}
\end{subequations}

With the $\gamma$ and $\Gamma$ parameters for a matter-dominated FRW
cosmological model, the unperturbed equation for $y_0$ has solutions
of the form
\begin{equation}
y_0 = -\frac{i}{\kappa\eta}
\frac{
 \left(1 + C\kappa\eta - \kappa^2\eta^2\right)\cos\left(\kappa\eta\right)
-\left(C - \kappa\eta - C\kappa^2\eta^2\right)\sin\left(\kappa\eta\right)
}{
\left(1+C\kappa\eta\right)\cos\left(\kappa\eta\right) -
\left(C-\kappa\eta \right)\sin\left(\kappa\eta\right)}, 
\end{equation}
where $C$ is a constant of the integration. In the limit of large $\kappa$ the evolution of $\kappa$ should decouple from the universal expansion; thus, we are led to choose $C=\pm i$, which eliminates the oscillatory terms in our general solution for $y_0$:
\begin{equation}
y_0 = 
\frac{\pm\kappa^3\eta^3-i}{\kappa\eta\left(1+\kappa^2\eta^2\right)}
\label{unpert-diff-eq}
\end{equation}
Consistent with our ansatz [cf. Eq. \eqref{eq:ansatz}] we choose the solution with positive $\Re(y_0)$: i.e., $C=+i$. 
Solving this equation for the phase $\phi'_0 = \kappa y_0$ we find
\begin{subequations}
\begin{eqnarray}
\Delta\phi_0(\eta) &=& \phi_0\left( 1 \right)-\phi_0\left( \eta \right)\\
 &=&
\left[\kappa\left(1-\eta\right) - \arctan\frac{\kappa\left(1-\eta\right)}{1+\kappa^2\eta}\right] 
-\frac{i}{2} \ln \left[\frac{1 + \kappa^2\eta^2}{\eta^2\left(1+\kappa^2\right)} \right]\\
&=&\left[
\frac{2}{\gamma}
\frac{\sqrt{1+z}-1}{\sqrt{1+z}}
- 
\arctan
\frac{
2\gamma\left(\sqrt{1+z}-1\right)
}{4+\gamma^2\sqrt{1+z}}
\right]
-
\frac{i}{2}
\ln\left[1+\frac{\gamma^2z}{4+\gamma^2}\right].
\end{eqnarray}
\end{subequations}
In the absence of the Chern-Simons correction an observer at rest with respect to slices of (cosmological) homogeneity will observe a passing gravitational plane-wave to undergo a change in phase $\Delta\phi_0(\eta)$ between cosmological time $\eta$ and the present epoch. 

With $y_0$  and equations~\eqref{eq:defY} we can evaluate the Chern-Simons contribution to the phase change owing to propagation from cosmological time $\eta_0$. Making use of the asymptotic
expansions for $y_{0,1}$ and $y_{1,0}$  [Eq.~\eqref{eq:asympt}] we find 
\begin{subequations}
\begin{eqnarray}
y_{0,1 (\text{R,L})} &\sim&
\frac{i\lambda_{\text{R,L}}}{2}
\left[
E(\eta)
-E(\eta_0)
e^{-2i\left[\Delta\phi_0(\eta_0)-\Delta\phi_0(\eta)\right]}
\right]
+\mathcal{O}(\gamma)\\
y_{1,0 (\text{R,L})} &\sim& 
-i\lambda_{\text{R,L}}
\left[
\Gamma(\eta)Z(\eta)-
\Gamma(\eta_0)Z(\eta_0)e^{-2i[\Delta\phi_0(\eta_0)-\Delta\phi_0(\eta)]}
\right]+\mathcal{O}(\gamma),
\end{eqnarray}
\end{subequations}
which may be integrated to find $\Delta\phi_{1 (\text{R,L})}$:
\begin{subequations}\label{eq:dPhi1}
\begin{eqnarray}
\Delta\phi_{1(\text{R,L})} &=& i\lambda_{\text{R,L}}\frac{\kappa}{2} \int_{\eta}^1 
\left[
\epsilon\frac{1}{\eta^4}\frac{\theta''(\eta)}{\theta''_0}-
2\gamma\zeta\frac{1}{\eta^5}\frac{\theta'(\eta)}{\theta'_0}
\right]\,d\eta + \mathcal{O}(\gamma),
\\
&=& i \lambda_{\text{R,L}} \frac{k_0}{H_0} \xi(\eta).
\end{eqnarray}
It is convenient to rewrite $\xi$ as a function of $z$:
\begin{equation}\label{eq:xi(z)}
\xi(z) = \alpha A(z) + \beta B(z),
\end{equation}
where
\begin{eqnarray}
\label{eq:A}
A(z) &=& \int_0^z dz\,\left(1+z\right)^{5/2}\frac{d\theta/dz}{(d\theta/dz)_0}\\
\label{eq:B}
B(z) &=& \int_0^z dz\,\left(1+z\right)^{7/2}\frac{d^2\theta/dz^2}{(d^2\theta/dz^2)_0}\\
\label{eq:alpha}
\alpha &=& -\gamma \zeta +
\frac{3 \epsilon}{2} \frac{\left(d\theta/dz\right)_0}{2\left(d^2\theta/dz^2\right)_0+3\left(d\theta/dz\right)_0},
\\
\label{eq:beta}
\beta &=& 
\frac{\epsilon\left(d^2\theta/dz^2\right)_0}{2(d^2\theta/dz^2)_0+3(d\theta/dz)_0},
\end{eqnarray}
\end{subequations}
and the subscript zero denotes present-day values of the subscripted quantities. The leading order Chern-Simons correction to the accumulated phase is thus pure imaginary, corresponding to an attenuation of one circular polarization state and an equal amplification of the other. The attenuation/amplification is linearly dependent on the wavenumber. The function $\xi(z)$ may be thought of as a ``form-factor'' that probes the past history of the coupling $\theta$. 

\subsection{Binary Inspiral at Cosmological Distances}\label{sec:binaries}

The proposed LISA gravitational wave detector is capable of observing coalescing  binary black hole systems at cosmological distances: for example, the gravitational waves associated with a pair of $10^6\,\text{M}_\odot$ black holes will be observable at redshifts $z$ approaching 30. Over the year leading up to the merger of two such black holes the binary's period will decrease by two orders of magnitude, leading to a corresponding decrease in the radiation wavelength and increase in the magnitude of the Chern-Simons correction. The time-dependent relationship between the radiation amplitude in the two polarization states thus carries with it the signature of Chern-Simons gravity and can be used to characterize the functional $\theta$ that describes the Chern-Simons correction to classical General Relativity. 

To calculate the signature left by the Chern-Simons correction on the gravitational waves from a coalescing binary system at redshift $z$, we begin with the radiation near the source. Treating, as before, the Chern-Simons correction as a perturbation, the quadrupole approximation to the radiation from the binary system in the neighborhood of the source is given by 
\begin{subequations}
\begin{eqnarray}
\widehat{\mathbf{h}} &=& \Re\left[\widehat{h}_{+} \; e_{+} + \widehat{h}_{\times} \; e_{\times}\right] \\
\widehat{h}_{+} &=& \frac{2\widehat{\mathcal{M}}}{d} \left[1 + \hat{\chi}^2\right]
\left[
  \widehat{\mathcal{M}} 
 \hat{k}(\hat{t})/2
  \right]^{2/3}
  \exp\left[-i\left(\hat\Phi(\hat{t}) - \hat{k}(\hat{t})d\right)\right],
\\
\widehat{h}_{\times} &=& \frac{4i\widehat{\mathcal{M}}}{d} \hat{\chi}\,
\left[
  \widehat{\mathcal{M}} 
  \hat{k}(\hat{t})/2
  \right]^{2/3} 
  \exp\left[-i\left(\hat\Phi(\hat{t}) - \hat{k}(\hat{t})d\right)\right],
\end{eqnarray}
\end{subequations}
where $d$ is the proper distance to the source and  
\begin{eqnarray}
\hat\Phi(\hat{t}) &=& - 2 \left(\frac{\hat{T} - \hat{t}}{5 {\widehat{\mathcal{M}}}}\right)^{5/8} + \hat{\delta},
\\
\hat{k}(\hat{t}) &=& \frac{2}{{\widehat{\mathcal{M}}}} \left( \frac{5}{256}
  \frac{{\widehat{\mathcal{M}}}}{\hat{T}-\hat{t}} \right)^{3/8}.
  \label{eq:k(t)}
\end{eqnarray}
The constants $\hat{T}$ and $\hat{\delta}$, which determine when coalescence occurs and the phase of the gravitational wave signal at some fiducial instant, are set by initial conditions. 
The quantities $\widehat{\mathcal{M}}$ and $\hat{\chi}$ are constants that depend on the binary system's component masses ($m_1$, $m_2$) and orientation with respect to the observer: 
\begin{subequations}
\begin{eqnarray}
\hat\chi &=& \left(
\begin{array}{l}
\text{cosine-angle between the orbital angular}\\
\text{momentum and the observer line-of-sight}
\end{array}
\right)\\
\widehat{\mathcal{M}} &=& \frac{m_1^{3/5} m_2^{3/5}}{\left(m_1 +
    m_2\right)^{1/5}} = \left(\text{``chirp'' mass}\right).
\end{eqnarray}
\end{subequations}
We ``hat'' all these quantities to remind us that, as expressed above, they are appropriate descriptions only in the neighborhood of the source where the Chern-Simons and cosmological corrections to the propagation of the waves may be neglected.

To describe the radiation after it has propagated to the detector we first describe the near-source radiation in terms of circular polarization states
\begin{subequations}\label{eq:nearSrc}
\begin{eqnarray}
\widehat{\mathbf{h}} &=& \Re\left[\widehat{h}_{R} \; e_{R} + \widehat{h}_{L} \; e_{L}\right],\\
\label{hr}
\widehat{h}_{R,L} &=& \sqrt{2} \frac{{\widehat{\mathcal{M}}}}{d} 
\left(\frac{\widehat{\mathcal{M}} k}{2}\right)^{2/3} 
\left(1 + \lambda_{R,L}\hat{\chi} \right)^2
  \exp\left[-i\left(\hat\Phi(\hat{t}) - \hat{k}(\hat{t})d\right)\right].
\end{eqnarray}
\end{subequations}
We are interested in the radiation incident on our detector today ($z=0$, $\eta=1$) from a source at redshift $z$. Matching the near-source description of the radiation [Eq.~\eqref{eq:nearSrc}) to our ansatz [Eq.~\eqref{eq:ansatz}] we find the description of the radiation after propagating to the detector from a redshift $z$:
\begin{subequations}
\begin{eqnarray}
\mathbf{h} &=& 
\Re\left[\widehat{h}_{R} \; e_{R} + \widehat{h}_{L} \; e_{L}\right],\\
h_{\text{R,L}} &=& \sqrt{2} \frac{\mathcal{M}}{d_L}
\left(\frac{\mathcal{M} k_0}{2}\right)^{2/3} 
\left(1 + \lambda_{R,L}\hat{\chi}\right)^2
\exp\left[-i
\left(
\Phi_0(t) - \kappa\left(1-\eta\right)
+\Delta\phi_0(t) + \Delta\phi_{1(\text{R,L})}(t)\right)\right]
\label{eq:hrl}
\end{eqnarray}
where
\begin{eqnarray}
\Phi_0(t) &=& - 2 \left(\frac{T - t}{5 \mathcal{M}}\right)^{5/8} + \delta,
\\
\label{k0}
k_0(t) &=& \frac{2}{\mathcal{M}} \left( \frac{5}{256}
  \frac{\mathcal{M}}{T-t}\right)^{3/8}\\
d_L &=& a_0\eta(1+z) = \left(\text{Luminosity distance to source}\right)\\
\mathcal{M} &=& \left(1+z\right)\widehat{\mathcal{M}}\\
\Delta\phi_0 &=&
\frac{2\left(\sqrt{1+z}-1\right)}{H_0\sqrt{1+z}}k_0(t)
- 
\arctan
2\gamma(t)
\frac{
\sqrt{1+z}-1
}{4+\sqrt{1+z}\gamma^2(t)}
\nonumber\\
&&\qquad{}
-
\frac{i}{2}
\ln\left[1+
\gamma^2(t)\frac{z}{4+\gamma^2(t)}\right]
\label{eq:phi0b}\\
\gamma(t)&=& \frac{H_0}{k_0(t)},
\end{eqnarray}
and $\Delta\phi_{1(\text{R,L})}$ given by equations \eqref{eq:dPhi1} above. 
\end{subequations}
Here $t$ is proper time as measured by a detector at rest with respect to the cosmological fluid at the present epoch ($\eta=1$), $k_0(t)$ is the instantaneous wavenumber of the wavefront passing the detector at observer time $t$, and $T$ and $\delta$ are, as before, constants of the integration. The correction $\Delta\phi_0$, which is the same for all polarizations, embodies $\mathcal{O}(k_0/H_0)$ corrections to the wave phase owing to the wave propagation through the time-dependent cosmological background. The correction $\Delta\phi_{1(\text{R,L})}$ is of opposite character for the two polarization states and embodies the (first-order) corrections to wave propagation owing to the Chern-Simons corrections to the Einstein Field Equations. 

Focus attention on the argument of the exponential in equation~\eqref{eq:hrl}. The term $\kappa\left(1-\eta\right)$ cancels the first term in equation~\eqref{eq:phi0b} for $\Delta\phi_0$, leading to 
\begin{equation}
h_{\text{R,L}} =
\sqrt{2} \frac{\mathcal{M}}{d_L}
\left(\frac{\mathcal{M} k_0}{2}\right)^{2/3} 
\left(1 + \lambda_{R,L}\hat{\chi}\right)^2
\exp
\left[-i
\left(
\Phi_0(t) - \frac{\gamma(t)}{2}\left(\sqrt{1+z}-1\right) + \Delta\phi_{1(\text{R,L})}(t)
\right)
\right]
\end{equation}
The observational effect of the Chern-Simons is readily identified by looking at the ratio of the polarization amplitudes $h_{R}$ and $h_{L}$:
\begin{eqnarray}
\frac{h_{\text{R}}}{h_{\text{L}}}
&=& \frac{\left(1+\hat{\chi}\right)}{\left(1-\hat{\chi}\right)}\exp\left[\frac{2k(t)\xi(z)}{H_0}\right]\\
&=& \frac{1+x}{1-x}
\end{eqnarray}
where $\xi$ is given by equations \eqref{eq:dPhi1} and $x$ may be interpreted as the \emph{apparent} inclination cosine-angle. The effect of the Chern-Simons correction on gravitational wave propagation is to confound the identification between polarization amplitude ratios and binary orbit inclination cosine-angle. \emph{In the same way that we say that the curvature of spacetime ``bends'' light passing close to strongly gravitating body we may say that the effect of the Chern-Simons correction is to ``rotate'' the apparent inclination angle of the binary system's orbital angular momentum axis either toward or away from us.} 

\section{Discussion}

\subsection{What can be measured?}\label{sec:measure}

Over the course of a year-long observation the LISA spacecraft constellation will measure the radiation in both polarizations of an incident gravitational wave train  associated with an inspiraling coalescing binary system. The relative amplitude of the two polarization will be determined by the orientation of the binary systems orbital plane to the observer line-of-sight and the form factor $\xi(z)$. A non-vanishing $\xi$ leads to a time-varying apparent inclination angle that, by nature of its time dependence, can (in principle) be measured directly from the apparent inclination angle's time variation.  

Other properties of an inspiraling binary can lead to an evolution of the (apparent) inclination cos-angle. Spin-orbit coupling leads to a real precession of the binary's orbital plane and a corresponding time-dependence in the actual  inclination cos-angle $\hat{\chi}$. Referring to  equation~\eqref{eq:hrl}, it is apparent that for small $|\hat{\chi}|\sim0$ an incremental change $\mu$ in $\hat{\chi}$ will lead to changes in $h_{\text{R,L}}$ that are indistinguishable from an increment in $x$ associated with $\xi$. Following Vecchio \cite[eq.\ 27--31]{vecchio:2004:loo} we note that, at first non-vanishing post-Newtonian order, spin-orbit interactions in an inspiraling binary system lead to 
\begin{equation}
\left(\frac{d\hat{\chi}}{dt}\right)_{\text{so}} \propto k_0^{2/3}(t).
\end{equation}
This is a different dependence on $k_0$ than the $\mathcal{O}(k_0)$ dependence associated with $\xi$. Thus, it remains in principle possible to distinguish the signature of Chern-Simons gravity in the signal from cosmologically distant coalescing binary black hole systems. The accuracy with which such a measurement can be made is the topic of the next subsection.

\subsection{How accurately can $\xi$ be measured?}

The most general astrophysical black hole binary system can be described by eleven independent parameters, which may be counted as 
two component masses; 
component spins and their orientation (six parameters);
orbital eccentricity; 
orbital phase; and a 
a reference time when the phase, spins and eccentricity are measured.
The gravitational wave signal in any particular polarization will depend on the description of the binary and six additional parameters that describe the binary's orientation with respect to the detector. These six additional parameters may be counted as 
orbital plane orientation (two angles);
source location with respect to the detector (distance and two position angles); 
orbit orientation in orbital plane (one angle) \footnote{This parameterization, while convenient for counting, is not the most appropriate for describing a coalescing binary, where spins, eccentricity, and orbital plane orientation all evolve with time.}.
To these seventeen parameters we now add $\xi$, which describes the effect of propagation through the birefringent Chern-Simons spacetime, for a total of eighteen parameters that are required to describe the signal from a coalescing binary system.

To-date, all analyses of expected parameter estimation errors have been made with under a set of approximations that focus attention on the measurement of component masses, source location (both distance and angular position), and the expected time of binary coalescence. Even the most sophisticated of these analysis ignore all but the leading-order contribution to the gravitational wave signal \emph{amplitude} at twice the orbital frequency\footnote{See \cite{arun:2007:hsh} for an analysis that relaxes this approximation} and assume that the orbital eccentricity is \emph{known} to vanish. These approximations are quite appropriate for their purpose (estimation of component masses, source location and expected time of coalescence); however, by ignoring all but the leading order contribution to the signal magnitude they are inadequate starting points for exploring the accuracy with which $\xi$, which affects only the signal amplitude in the different polarizations, can be bounded \footnote{This applies also to the assumption of zero eccentricity: if, as is likely, the estimation errors associated with eccentricity and $\xi$ are correlated it will be necessary to include an eccentricity-like parameter in the analysis in order to avoid under-estimating the errors associated with the measurement of $\xi$.}. Evaluating and presenting the errors associated with the measurement of $\xi$ via a full co-variance matrix analysis is thus a formidable enterprise, to be addressed in a future work. 

Nevertheless, through a series of plausible approximations it is possible to make a crude estimate of the accuracy with which $\xi$ can be determined. To begin, assume we have two gravitational wave detectors such that, via a linear combination of observations made at each, we can can synthesize two other detectors with one exclusively sensitive to $h_{\text{R}}$ and and one exclusively sensitive to $h_{\text{L}}$. Write the scalar detector response of each of these detectors as
\begin{equation}
m_{\text{R,L}}(t) = \exp\left[\mu_{\text{R,L}}(t) + i\psi_{\text{R,L}}(t)\right],
\end{equation}
for real $\mu_{\text{R,L}}$ and $\psi_{\text{R,L}}$. Next, note that the parameters that describe a coalescing binary system can be divided into two groups: those that principally affect only the signal amplitude (i.e., $\mu(t)$) and those that affect only or principally the real part of the signal phase (i.e., $\psi(t)$). The first group includes 
distance,  
source orientation with respect to the observer line-of-sight, and
$\xi$. The second group includes the orbital phase, sky location
(through its affect on the Doppler correction to the signal phase as
the detector orbits about the sun), the instantaneous binary period at
some fiducial moment, and the parameters associated with spin and
orbital angular momentum \footnote{This approximation is weakest for
  the spin and orbital angular momentum parameters: see
  \cite{van-den-broeck:2006:bbh}.}. If we approximate each detectors'
noise as white with one-sided noise power spectral density $S_{0}$ then the elements of the inverse covariance matrix $\boldsymbol{\Gamma}$ --- the so-called Fisher matrix --- are given by \cite{finn:1992:dma,finn:1993:obi}
\begin{equation}
\Gamma_{ij} = \sum_{k=\text{R,L}}\frac{2}{S_{0}} \int_{t_i}^{t_f}
\Re\left(\frac{\partial{m_k}}{\partial
    x^i}\right)\,\Re\left(\frac{\partial{m_k}}{\partial x^j}\right)\, dt
\end{equation}
where the integration is over the observation period $(t_i,t_f)$ and the $x^i$ are the parameters that characterize the incident gravitational wave, which we have divided into two groups. Matrix elements $\Gamma_{ij}$ where $x^i$ and $x^j$ belong to different groups will be much smaller than elements where $x^i$ and $x^j$ belong to the same group. Setting the cross-group elements to zero we obtain an approximate $\boldsymbol{\Gamma}$ that is block diagonal, with one block corresponding to $\Gamma_{ij}$ with $(x^i,x^j)$ drawn from the first group and the other block corresponding to $\Gamma_{ij}$ with $(x^i,x^j)$ drawn from the second group. Estimation uncertainties of parameters in either group can now be determined independently of the parameters in the other group. 

Focus attention now on those parameters that affect only $\mu(t)$, the signals amplitude evolution. The leading order dependence of the amplitude $|h_{\text{R,L}}|$ on the binary systems parameters is given by 
\begin{equation}
A_{\text{R,L}} = 
|h_{\text{R,L}}| = 
\left(1+\lambda_{\text{R,L}}\hat\chi_0\right)^2\frac{2\mathcal{M}}{d_L}\left[\frac{k_0(t)\mathcal{M}}{2}\right]^{2/3}
\exp\left(\lambda_{\text{R,L}}\xi \frac{k_0(t)}{H_0}\right)
\end{equation}
where $\mathcal{M}$ is assumed known. Setting aside the antenna pattern factors associated with the projection of the signal onto the LISA detector (which depend only on the known source sky position and the LISA orbital ephemeris), assuming that there is no real precession in the binary system under observation (i.e., $\hat{\chi}_0$),  and that $k(t)$ is given by equation \eqref{eq:k(t)} the inverse of the covariance matrix --- the so-called Fisher matrix, $\boldsymbol{\Gamma}$ --- associated with the amplitude measurements is a symmetric $3\times3$ matrix with elements
\begin{subequations}
\begin{eqnarray}
\Gamma_{\mathcal{D}\mathcal{D}} &=& \frac{1}{S_0} \int_{t_i}^{t_f}
\left(A_{\text{R}}^2 + A_{\text{L}}^2 \right)\; dt
\simeq
8\left(1+6\hat\chi_0^2+\hat\chi^4_0\right)\left(\frac{\mathcal{M}}{d_L}\right)^2\mathcal{I} + \mathcal{O}(\xi)\\
\Gamma_{\mathcal{D}\hat\chi_0} &=& \frac{1}{S_0}  \int_{t_i}^{t_f}
\left[ \frac{2}{1-\hat\chi_0}A_{\text{L}}^2 
-\frac{2}{1+\hat\chi_0}A_{\text{R}}^2 \right] dt  
\simeq
-16\hat\chi_0\left(3+\hat\chi_0^2\right)\left(\frac{\mathcal{M}}{d_L}\right)^2\mathcal{I} + \mathcal{O}(\xi)\\
\Gamma_{\mathcal{D}\xi} &=& \frac{1}{S_0} \int_{t_i}^{t_f}\frac{k_0(t)}{H_0}
\left( A_{\text{L}}^2 - A_{\text{R}}^2 \right) dt  
\simeq
-\frac{64\hat\chi_0\left(1+\hat\chi_0^2\right)}{\mathcal{M}H_0}\left(\frac{\mathcal{M}}{d_L}\right)^2\mathcal{J} + \mathcal{O}(\xi)\\
\Gamma_{\hat\chi_0\hat\chi_0} &=& \frac{1}{S_0}  \int_{t_i}^{t_f}
\left[ 
\frac{4 A_{\text{R}}^2}{\left(1 + \hat{\chi}_0\right)^2}
+ 
\frac{4 A_{\text{L}}^2}{\left(1 - \hat{\chi}_0\right)^2}
\right]
 dt
\simeq
32\left(1+\hat\chi_0^2\right)\left(\frac{\mathcal{M}}{d_L}\right)^2\mathcal{I} + \mathcal{O}(\xi)\\ 
\Gamma_{\hat\chi_0\xi} &=& \frac{1}{S_0} \int_{t_i}^{t_f}
\frac{2 k_0(t)}{H_0} 
\left[
\frac{A_{\text{R}}^2}{1 + \hat{\chi}_0} +
\frac{A_{\text{L}}^2}{1 - \hat{\chi}_0}
\right]
dt 
\simeq
\frac{32 \left(1+3\hat\chi^2_0\right)}{\mathcal{M}H_0}\left(\frac{\mathcal{M}}{d_L}\right)^2\mathcal{J} + \mathcal{O}(\xi)\\ 
\Gamma_{\xi\xi} &=& \frac{1}{S_0}  \int_{t_i}^{t_f}\frac{k_0^2(t)}{H_0^2}
\left(A_{\text{R}}^2 + A_{\text{L}}^2  \right) dt 
\simeq
\frac{32\left(1+6\hat\chi_0^2+\hat\chi^4_0\right)}{\left(\mathcal{M}H_0\right)^2}\left(\frac{\mathcal{M}}{d_L}\right)^2
\mathcal{K} + \mathcal{O}(\xi)
\end{eqnarray}
where
\begin{eqnarray}
\mathcal{D} &=& \ln d_L\\
\mathcal{I} &=& \int_{t_i}^{t_f} \left(\frac{k_0(t)\mathcal{M}}{2}\right)^{4/3}\frac{dt}{S_0}  
=\frac{5}{192}\int_{k_{\min}}^{k_{\max}}\left(\frac{k\mathcal{M}}{2}\right)^{-7/3}
\frac{\mathcal{M}^2 dk}{S_0}
=-\frac{2^{1/3}\mathcal{M}}{S_0}\frac{5}{64}\left.\left(k\mathcal{M}\right)^{-4/3}\right|_{k_{\min}}^{k_{\max}}
\\
\mathcal{J} &=& \int_{t_i}^{t_f} \left(\frac{k_0(t)\mathcal{M}}{2}\right)^{7/3}\frac{dt}{S_0} 
=\frac{5}{192}\int_{k_{\min}}^{k_{\max}}\left(\frac{k\mathcal{M}}{2}\right)^{-4/3}
\frac{\mathcal{M}^2 dk}{S_0}
=-\frac{2^{1/3}\mathcal{M}}{S_0}\frac{5}{32}\left.\left(k\mathcal{M}\right)^{-1/3}\right|_{k_{\min}}^{k_{\max}}\\
\mathcal{K} &=& \int_{t_i}^{t_f} \left(\frac{k_0(t)\mathcal{M}}{2}\right)^{10/3}\frac{dt}{S_0}
=\frac{5}{192}\int_{k_{\min}}^{k_{\max}}\left(\frac{k\mathcal{M}}{2}\right)^{-1/3}
\frac{\mathcal{M}^2 dk}{S_0}
=\frac{2^{1/3}\mathcal{M}}{S_0}\frac{5}{128}\left.\left(k\mathcal{M}\right)^{2/3}\right|_{k_{\min}}^{k_{\max}}
.
\end{eqnarray}
\end{subequations}
and we have taken advantage of the fact that for inspiraling compact binary systems in the quadrupole approximation $k(t)$ is monotonic in $t$ to re-express the integrals over the interval $(t_i,t_f)$ as integrals over $[k(t_i),k(t_f)]=(k_{\min},k_{\max})$. 
In the particular case of a binary seen plane-on ($\hat\chi_0=0$), the
$(\mathcal{D}\hat\chi_0)$  and $(\mathcal{D}\xi)$ blocks of $\boldsymbol{\Gamma}$ are diagonal, leading to 
\begin{subequations}
\begin{eqnarray}
\nu_{\hat\chi_0\hat\chi_0} &=& \frac{1}{4\rho^2} \frac{\mathcal{K}
  \mathcal{I}}{\mathcal{I} \mathcal{K} - \mathcal{J}^2} 
\\
\nu_{{\xi\xi}} &=& \frac{\left(\mathcal{M}H_0\right)^2}{4\rho^2}
\frac{\mathcal{I}^2}{\mathcal{I} \mathcal{K} - \mathcal{J}^2} 
\end{eqnarray}
where $\nu_{ij}$ is the ensemble average co-variance 
\begin{equation}
\nu_{ij} = \overline{
\left(x^i-\overline{x_i}\right)
\left(x^j-\overline{x_j}\right)
} = \left(\boldsymbol{\Gamma}^{-1}\right)_{ij}
\end{equation}
and we have expressed the $\nu_{ij}$ in terms of the ensemble average amplitude-squared signal-to-noise ratio $\rho^2$
\begin{equation}
\rho^2 =\frac{1}{S_0} \int_{t_i}^{t_f}dt\,\left(A_R^2 + A_L^2\right)
\end{equation}
\end{subequations}

Focus attention on a binary system of two black holes at redshift $z$, each with mass $M=10^6\,M_{\odot}\left(1+z\right)^{-1}$. Over the final year before coalescence the radiation wavelength $2\pi/k$ observed at the detector will range from $c\left(10^{-4}\,\text{Hz}\right)^{-1}$ to $c\left(10^{-2}\,\text{Hz}\right)^{-1}$. For such a system, 
\begin{eqnarray}
\rho &=&  \frac{10^5h_{100}}{1+z-\sqrt{1+z}} \left(\frac{10^{-40}\,\text{Hz}^{-1}}{S_0}\right)^{1/2}
\\
\nu_{\xi\xi}&=& 3.1\times10^{-40}\left(\frac{S_0}{10^{-40}\,\text{Hz}^{-1}}\right)\left(1+z-\sqrt{1+z}\right)^2
\end{eqnarray}
Observation of binary systems like these at $z=15$ by LISA will be capable of placing 
a ``1-sigma'' upper bound on $\xi$ of order
$10^{-19}$.

\subsection{How large might $\xi$ be?}
\label{models}

To estimate $\xi$ [cf.~Eq.~\eqref{eq:xi(z)}] we must invoke a
theoretical model for the functional $\theta[\phi(z)]$.  As described
in the introduction, perturbative string theory requires a
Chern-Simons correction to the Einstein-Hilbert action \footnote{Other
  theories, such as loop quantum gravity, may also possess a similar
  correction in the low-energy limit; however, no effective
  Hamiltonian has yet been found in the full theory of loop quantum
  gravity and cosmology, which prevents further speculation as to
  whether such an extension of general relativity requires a
  Chern-Simons modification.}. Here we describe a different mechanism, that can also lead to the
presence of a Chern-Simons correction. Consider the
back-reaction of a $\mathcal{N}\rm=1$ supersymmetric Yang-Mills theory
in a curved background (cf.~\cite[Appendix A]{alexander:2007:glf}) with action
\begin{equation}
\label{String-Chern-Simons}
S_{\text{CS}} = \frac{1}{16 \pi} \int d^4 x  \; \mathcal{F}\rm_{1}(S) \; 
\left(R \, {}^{\star}R \right),
\end{equation}
where $S$ is the glueball superfield and $\mathcal{F}\rm_{1}(S)$,
which plays the role of $\theta$ in equation~\eqref{eq:CS-action}, can
be exactly evaluated by using perturbative matrix model technology
developed in~\cite{dijkgraaf:2004:mma}. Within this Yang-Mills framework, $\theta$ is a functional $\theta[\varphi]$ of some pseudo-scalar field $\varphi$, the gravitational axion, that depends only on conformal time \cite{alexander:2005:bgw}. The functional $\theta[\varphi]$ can be expressed as
\begin{equation}\label{f}
\theta[\varphi] = \frac{\mathcal{N}\ell^2_{\text{Pl}}}{2 \pi}
\frac{\varphi}{M_{\text{Pl}}}, 
\end{equation}
where $\ell_{\text{Pl}}=\left(\hbar G/c^3\right)^{1/2}$ is the Planck length, $M_{\text{Pl}} =
(\ell_{\text{Pl}}\sqrt{8 \pi})^{-1}$ is the reduced Planck mass and
${\mathcal{N}}$ is a dimensionless constant. Through use of the  low-energy effective
$4$-dimensional heterotic string action 
the constant $\mathcal{N}$ can be evaluated 
in terms of the
ten-dimensional fundamental string energy scale $M_{\text{s}}$ and
the dimensionless string coupling $g_{\text{s}}$,
\begin{equation} 
  {\mathcal{N}} = \left(\frac{M_{{\text{Pl}}}}{M_{\text{s}}}\right)^{2}
  \frac{1}{\sqrt{g_{\text{s}}}},
\end{equation}
leading to 
\begin{equation}\label{eq:f}
\theta[\varphi] = \frac{1}{16\pi^2 M_{\text{Pl}}^2}
\frac{\varphi}{M_{\text{s}}}
\frac{M_{\text{Pl}}}{M_{\text{s}}\sqrt{g}_{\text{s}}}. 
\end{equation}
Assuming that $\varphi$, which has units of inverse length, evolves
with the Hubble parameter $H\propto \eta^{-3}$ we have
\begin{equation}
A(z) = B(z) = -\frac{2}{11}\left[\left(1+z\right)^{11/2}-1\right]
\end{equation}
and
\begin{equation}
\xi = -\frac{1}{11}\left[\left(1+z\right)^{11/2}-1\right]
\left(\epsilon - 2 \gamma \zeta \right) 
\end{equation}
with $(\epsilon - \gamma \zeta)$ of the order of 
\begin{subequations}
\begin{eqnarray}
\epsilon - 2 \gamma \zeta \simeq \frac{\left(1.8+3.5h_{100}^2\right)\times10^{-120}}{\sqrt{g_{\text{s}}}}
\left(\frac{\varphi_0}{M_{\text{s}}}\right)
\left(\frac{10^{16}\,\textrm{GeV}}{M_{\text{s}}}\right).
\end{eqnarray}
\end{subequations}
The size of $\xi$ thus depends on the present value of the field $\varphi_0$, the fundamental string energy scale $M_s$ and the string coupling $g_{\text{s}}$, none of which are constrained by present-day theory. 

The lesson to draw from the discussion of this scenario is that the
magnitude of any Chern-Simons correction depends strongly on the
external theoretical framework that prescribes the functional
$\theta[\varphi]$. For
non-vanishing string coupling in the perturbative string theory scenario the Chern-Simons correction seems
undetectable owing to the Planck scale suppression of the decay
constant of the universal gravitational axion field $\varphi$. 
However, this model and the associated expected scale of $\epsilon-2\gamma\zeta$ applies only to the \emph{perturbative} sector of
string theory and, in particular, when Ramond-Ramond charges are
turned off.  If present these additional degrees of freedom do couple and
source the Chern-Simons correction, leading to a larger decay constant
(e.g, D3 branes always excite the Chern-Simons interaction
in four dimensions). In a recent work, Svrcek \cite{svrcek:2006:ais} noted
that, due to non-perturbative gravitational instanton corrections, the
Chern-Simons coupling in the non-perturbative sector is currently
incalculable.  Even within the perturbative
framework there are theoretical frameworks where
$\xi$ could become significant: e.g., if the string coupling $g_{\text{s}}$
vanishes at late times
\cite{brandenberger:1989:sie,%
tseytlin:1992:eos,%
nayeri:2006:pss,%
sun:2006:ccm,%
wesley:2005:cct,%
alexander:2000:bgi,%
brandenberger:2002:lpi,%
battefeld:2006:sgc,%
brandenberger:2006:sgc,%
brandenberger:2007:sgc,%
brax:2004:bwc}. Therefore, within the full string theory framework, a
larger coupling, which would push the stringy
Chern-Simons correction into the observational window, is not excluded and bounding it
places a constraint on string theory motivated corrections to classical general relativity. 

Setting aside string theory for a moment, other theoretical frameworks
lead to a Chern-Simons correction to the Einstein-Hilbert action. Jackiw and
Pi~\cite{jackiw:2003:cmo} showed that the embedding of the
$3$-dimensional Chern-Simons action into $4$-dimensional General
Relativity leads to Chern-Simons gravity as described by
equation~\eqref{eq:CS-action}. Quantum mechanically, the Chern-Simons
correction encodes information about the Immirzi parameter in loop
quantum gravity and is also related to torsion. In this case, the Chern-Simons
term is \emph{required} to ensure invariance under large gauge
transformations~\cite{ashtekar:1989:cpi}. In both these classical and
quantum mechanical scenarios the coupling of the Chern-Simons term to
the Einstein-Hilbert action is theoretically unconstrained and is
consistent with a coupling of order unity.  Observations of gravitational waves from inspiraling binary black hole systems can thus be used to probe the strong gravity sector of quantum gravity. 

\section{Conclusions}\label{sec:conclusions}

Chern-Simons type corrections to the Einstein-Hilbert action are strongly motivated by string theory, loop quantum gravity, and other scenarios. In all cases these corrections lead to an amplitude birefringence for gravitational wave propagating through space time. We have evaluated the correction to the gravitational wave amplitude for waves propagating over cosmological distances in a matter-dominated Friedmann-Robertson-Walker cosmology. In the case of the gravitational waves from inspiraling binary black hole systems the effect of the spacetime birefringence is an \emph{apparent} time-dependent change in the inclination angle between the binary system's orbital angular  momentum and the line-of-sight to the detector. (This change is apparent in the same sense that light is ``bent'' upon passage nearby a strongly gravitating object.) Sufficiently long observations of a binary system will enable this apparent rotation to be distinguished from the real rotation caused by spin-orbit and spin-spin angular momentum interactions in the binary system. Observations of just this kind will be possible using the LISA gravitational wave detector \cite{merkowitz:2007:lis,baker:2007:lpt,nrc:2007:nbe}, which will be able to observe the inspiral of massive black hole binaries at redshifts approaching 30 for periods of a year or more. Gravitational wave observations of these systems with LISA may thus provide the first test of string theory or other quantum theories of gravity: yet another way in which gravitational wave observations can act a unique tool for probing the fundamental nature of the universe. 

\acknowledgments We are grateful to S. James Gates Jr. for encouraging
us to look into the possibility of testing string theories with the
Green-Schwarz term. We also thank Lubos Motl, Benjamin Owen, Michael
Peskin, Andy Strominger, Emanuele Berti, Clifford Will, Carlos
Sopuerta, Eric Poisson and Scott Hughes for enlightening discussions
and comments.

SA acknowledges the support of the Eberly College of Science.  NY
acknowledges support from National Science Foundation awards PHY
05-55628 and PHY 02-45649. LSF acknowledges the support of NSF awards
PHY 06-53462 and PHY 05-55615, and NASA award NNG05GF71G. Lastly, we
acknowledge the support of the Center for Gravitational Wave Physics,
which is funded by the National Science Foundation under Cooperative
Agreement PHY 01-14375.


\end{document}